\documentclass[letterpaper,english,aps,prl,groupedaddress,superscriptaddress,floatfix,twocolumn,showpacs,amsfonts,amssymb]{revtex4}
\usepackage[T1]{fontenc}
\usepackage[latin1]{inputenc}
\usepackage{graphicx}
\usepackage{amssymb}
\usepackage{overpic}
\usepackage{babel}

\newcommand{\beq}{\begin{equation}}
\newcommand{\eeq}{\end{equation}}
\newcommand{\bea}{\begin{eqnarray}}
\newcommand{\eea}{\end{eqnarray}}

\begin{document}
\title{Retardation effects in the Holstein-Hubbard chain at half-filling}

\author{Ka-Ming Tam}
\affiliation {Department of Physics, Boston University, 590 Commonwealth
  Ave., Boston, MA 02215} 
\author{S.-W. Tsai}
\affiliation{Department of Physics, University of California, Riverside, CA
  92508} 
\author{D.~K.~Campbell}
\affiliation {Department of Physics, Boston University, 590 Commonwealth
  Ave., Boston, MA 02215} 
\author{A.~H.~Castro Neto}
\affiliation {Department of Physics, Boston University, 590 Commonwealth
  Ave., Boston, MA 02215}

\date{\today}

\begin{abstract}

The ground state phase diagram of the half-filled one-dimensional Holstein-Hubbard model 
contains a charge-density-wave (CDW) phase, driven by the electron-phonon (e-ph) coupling,
and a spin-density-wave (SDW) phase, driven by the on-site electron-electron (e-e)
repulsion. Recently, the existence of a third phase, which
is metallic and lies in a finite region of parameter space between these two gapped phases,
has been claimed. We
study this claim using a renormalization-group method for interacting
electrons that has been extended to include also e-ph couplings. Our method [\onlinecite{Tsai}] 
treats e-e and e-ph interactions on an equal footing and takes retardation
effects fully into account. We find a direct transition between the spin- and
charge-density wave states. We study the effects of retardation, which are
particularly important near the transition, and find that Umklapp processes {\it
at finite frequencies} drive the CDW instability close to
the transition. We also perform determinantal quantum Monte Carlo
calculations of correlation functions to confirm our results for the phase diagram.

\end{abstract}

\maketitle

The interplay between electron-electron (e-e) and electron-phonon (e-ph) interactions leads to 
important effects in low-dimensional materials 
such as molecular crystals, charge transfer solids \cite{Organics}, conducting polymers \cite{Polymers}, and 
fullerenes \cite{Fullerene}.  In narrow band electronic materials, perhaps the
simplest model capturing this interplay is the Holstein-Hubbard model (HHM), where the  
e-e interactions are described by a on-site repulsive Coulomb term, and the electrons are coupled to 
dispersionless optical phonons in localized vibrational modes \cite{Holstein}.  

In the one-dimensional HHM (1DHHM) at half-filling, early quantum Monte Carlo (QMC) calculations \cite{Hirsch} suggested that there
are only two phases: the Peierls charge-density-wave (CDW) 
and the Mott spin-density-wave (SDW) state. The boundary between these two phases was predicted to lie along the
line in parameter space where an ``effective'' e-e interaction vanishes: 
$U_{\rm eff} = U - 2 g_{\rm ep}^2/\omega_{0} \simeq 0$, where $U$ is the Hubbard on-site e-e
repulsion, $g_{\rm ep}$ is the electron-phonon coupling, and $\omega_0$ is the phonon frequency. 
More recently, several authors have proposed that a third phase might exist near $U_{\rm eff} \simeq 0$: a
metallic, Luttinger liquid, phase \cite{Wu,Jeckelmann,Takada}, 
or an off-site pairing superconducting phase 
\cite{Takada2}. Large scale QMC studies \cite{Clay} have indicated that there is a metallic region with 
dominant superconducting (SC) pairing correlations between the CDW and SDW regions. DMRG studies \cite{Tezuka} suggest that 
SC does not exist but instead that both the spin and charge gaps vanish only for
$U_{\rm eff} \simeq 0$, suggesting that a metallic phase (with no dominant SC
correlations) may exist only exactly on the boundary between the CDW and SDW phases.  
This is also the conclusion of two-step renormalization-group studies
\cite{Bindloss} and Lanczos diagonalization \cite{Feshke}. To attempt to determine which
of these scenarios is correct, we study the problem here using a recently developed extended 
renormalization group approach \cite{Tsai}.   

At half-filling, Umklapp scattering creates a strong tendency to open a charge gap. From the perspective of
weak-coupling approaches, it is highly non-trivial to have a finite metallic, or SC, region. If such a phase is to exist, 
it must be that the dynamical nature of the phonons effectively suppresses Umklapp scattering. Therefore, 
retardation effects must be taken into account in order to investigate this issue. For this purpose, 
we use a multiscale functional renormalization-group (MFRG) method \cite{Tsai}. 
Our MFRG is an extension of the RG for interacting fermions \cite{Shankar} that are also coupled to bosonic modes and 
applies to both weak ($\lambda \ll 1$) and strong ($\lambda \gg 1$) electron-phonon coupling limit 
($\lambda = 2 N(0) g_{\rm ep}^2/\omega_0$, $N(0)$ is the electron density of states at the Fermi level). 
For a spherical Fermi surface, the MFRG reproduces Eliashberg's theory at the SC instability \cite{Tsai}, 
and it has also been applied in the study of effects of phonons in ladder systems \cite{Tam}.

The 1DHHM is given by the Hamiltonian 
\begin{eqnarray}
H &=& - t \sum_{i,\sigma}(c_{i+1,\sigma}^{\dagger}c_{i,\sigma}
    + H.c.) + U\sum_{i}n_{i,\uparrow}n_{i,\downarrow}\nonumber\\
    & & + g_{\rm ep} \sum_{i,\sigma}(a_{i}^{\dagger}+a_{i})n_{i,\sigma} +
    \omega_{0} \sum_{i} a_{i}^{\dagger}a_{i}, 
\end{eqnarray}
where $c_{i,\sigma}^{\dagger}$ ($c_{i,\sigma}$) is an electron creation
(annihilation) operators at site $i$ with spin $\sigma$, $n_{i\sigma}$ is the electron number operator, 
$a_{i}^{\dagger}$ ($a_{i}$) is a creation
(annihilation) operator for an optical phonon at site $i$, $t$ is the
nearest-neighbor electron hopping integral. We use units such that
$t=1=\hbar$.

Using a path integral formulation and integrating out the phonon
fields exactly, we find that the effective (retarded) e-e interaction becomes \cite{Tsai}:
\begin{eqnarray}
g(\underline{k}_1,\underline{k}_2,\underline{k}_3,\underline{k}_4) = U -
\frac{2g_{\rm ep}^2\omega_{0}}{[\omega_{0}^2 +(\omega_1 - \omega_4)^2]}, 
\label{eq:bare}
\end{eqnarray}
where $\underline{k} = (k,\omega)$. We use a notation in which, after
scattering, an incoming electron with momentum and frequency $\underline{k}_1$
($\underline{k}_2$) goes out with $\underline{k}_4$ ($\underline{k}_3$), so
that $\underline{k}_1 + \underline{k}_2 = \underline{k}_3 + \underline{k}_4$. 
In the anti-adiabatic limit, where
$\omega_0 \rightarrow \infty$, all the electronic frequency dependences are suppressed,
and the HHM maps onto the standard Hubbard model
with a renormalized $U_{\rm eff}$. At half-filling, its ground state is
charge-gapped SDW for repulsive interactions and spin-gapped degenerate
CDW/SC for attractive interactions. The transition between SDW and
degenerate CDW/SC occurs when the bare coupling changes sign, that is when
$U_{\rm eff}=0$.

\begin{figure}[bth]
\centerline{
\includegraphics*[height=0.23 \textheight,width=0.36\textwidth,viewport=80
150 500 555,clip]{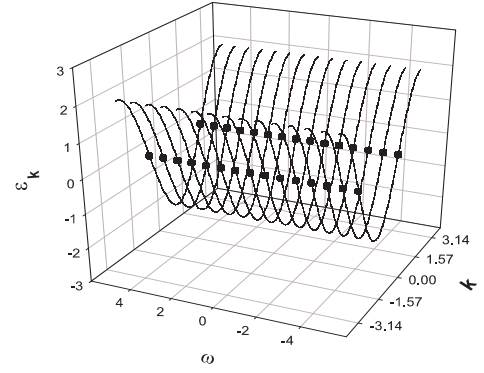}} \caption{Discretization of the momenta
in the Brillouin zone and frequencies in the frequencies axis. This figure shows the case $N_k=2, N_w =15$.}
\label{patches}
\end{figure}

In the MFRG approach at the one-loop level, the RG flow equations for the coupling {\it functions}, 
$g(\underline{k}_1,\underline{k}_2,\underline{k}_3,\underline{k}_4)$ with initial conditions 
given by (\ref{eq:bare}), are given by \cite{Tsai}:
\begin{eqnarray}
&&\frac{dg(\underline{k}_1,\underline{k}_2,\underline{k}_3)}{d \Lambda} =
\nonumber\\
&-&\!\!\!\!\int\!d\underline{p} \frac{d}{d\Lambda}
[G_{\Lambda}(\underline{p})G_{\Lambda}(\underline{k})]
g(\underline{k}_1,\underline{k}_2,\underline{k}) 
g(\underline{p},\underline{k},\underline{k}_3)
\nonumber\\
&-&\!\!\!\!\int\!d\underline{p} \frac{d}{d\Lambda}
 [G_{\Lambda}(\underline{p})G_{\Lambda}(\underline{q}_1)]
g(\underline{p},\underline{k}_2,\underline{q}_1)
g(\underline{k}_1,\underline{q}_1,\underline{k}_3) 
\nonumber\\
&-&\!\!\!\!\int\!d\underline{p} \frac{d}{d\Lambda}
 [G_{\Lambda}(\underline{p}) G_{\Lambda}(\underline{q}_2)]
 [-\!2g(\underline{k}_1,\underline{p},  
 \underline{q}_2)g(\underline{q}_2,\underline{k}_2,\underline{k}_3)
\nonumber\\
&+&\!\!g(\underline{p},\underline{k}_1,\underline{q}_2)
g(\underline{q}_2,\underline{k}_2,\underline{k}_3)\!+
\!g(\underline{k}_1,\underline{p},\underline{q}_2) 
g(\underline{k}_2,\underline{q}_2,\underline{k}_3)],
\label{eq:rg1}
\end{eqnarray}
where $\underline{k}=\underline{k}_1+\underline{k}_2-\underline{p}$,
$\underline{q}_1=\underline{p}+\underline{k}_3-\underline{k}_1$,
$\underline{q}_2=\underline{p}+\underline{k}_3-\underline{k}_2$, 
$\int d\underline{p}=\int
dp\sum_{\omega}1/(2\pi\beta)$, and $G_{\Lambda}$ is the self-energy corrected
propagator at energy cut-off $\Lambda$.
Since the interaction vertices are frequency dependent, there are also
self-energy corrections. At the one-loop level, the self-energy MFRG equation is:
\begin{eqnarray}
\frac{d \Sigma(\underline{k})}{d \Lambda} = 
&-&\!\!\!\!\int\!d\underline{p} \frac{d}{d\Lambda} [G_{\Lambda}(\underline{p})]
[2g(\underline{p},\underline{k}, \underline{k}) -
g(\underline{k},\underline{p},\underline{k})]. 
\label{eq:rg2}
\end{eqnarray}
We have solved the coupled integral-differential equations,
(\ref{eq:rg1}) and (\ref{eq:rg2}), numerically with two
Fermi points ($N_k = 2$) and by dividing the frequency axis into fifteen segments ($N_\omega = 15$). 
Fig. \ref{patches} shows the discretization scheme for $N_k = 2$ and $N_\omega = 15$.

We next calculate within our MFRG approach the RG flow of susceptibilities in the  
static (zero frequency) and long-wavelength limit.  
In particular, the SC susceptibility is given by:
$\chi^{{\rm SC}}_{\Lambda}(0,0)\!\!\!=\!\!\!\int D(1,2)
\langle c_{p_{1},\downarrow}c_{-p_{1},\uparrow}
c_{-p_{2},\uparrow}^{\dagger}c_{p_{2},\downarrow}^{\dagger}\!\rangle$; and the SDW and CDW susceptibilities can be written as:
$\chi^{\delta}_{\Lambda}(\pi,0)\!\!=\!\!\!\int D(1,2)
\langle
c_{p_{1},\sigma_{1}}^{\dagger}c_{p_{1}\!+\!\pi,\sigma_{1}}
c_{p_{2}\!+\!\pi,\sigma_{2}}^{\dagger}c_{p_{2},\sigma_{2}} \rangle $, 
where $p_i$ is the momentum at energy $\xi_{i}$,
$\int D(1,2)\equiv \int_{|\xi_{1}|>\Lambda}d\xi_{1}J(\xi_{1})
\int_{|\xi_{2}|>\Lambda}d\xi_{2}J(\xi_{2})
\sum_{\sigma_{1},\sigma_{2}}s_{\sigma_{1}}s_{\sigma_{2}}$, 
and $J(\xi)$ is the Jacobian for the coordinate transformation
from $k$ to $\xi_k$. For $\delta = {\rm SDW}$: $s_\uparrow =1, s_\downarrow =-1$, and for $\delta = {\rm CDW}$: $s_\uparrow =1, 
s_\downarrow = 1$.     The dominant instability is determined by the most divergent susceptibility as the cut-off $\Lambda$ is 
lowered. The RG flow for the SC susceptibility is given by:
\begin{eqnarray} 
\label{SC_susceptibility RGE}
\!\!\!\!\!\!\!\!\frac{d \chi^{{\rm SC}}_{\Lambda}\!\!(0,\!0)}{d\Lambda}\!\!\!&=&
\!\!\!\int\!\!\!d\underline{p} \frac{d}{d\Lambda}[
G_{\Lambda}(\underline{p})G_{\Lambda}(-\underline{p})]
(Z^{{\rm SC}}_{\Lambda}\!(\underline{p}))^2, \\
\!\!\!\!\!\!\!\!\frac{d Z^{{\rm SC}}_{\Lambda}\!(\underline{p})}{d\Lambda} \!\!\!\!&=&\!\!\!
-\!\!\!\int\!\!\!d\underline{p}^{\prime}\!\frac{d}{d\Lambda}[
G_{\Lambda}(\underline{p}^{\prime})G_{\Lambda}(\!-\underline{p}^{\prime})]
Z^{{\rm SC}}_{\Lambda}\!(\underline{p^{\prime}})
g^{{\rm SC}}\!(\underline{p^{\prime}},\underline{p}), 
\end{eqnarray}
where $g^{{\rm SC}}(\underline{p^{\prime}},\underline{p}) =
g(\underline{p^{\prime}}\!,-\underline{p^{\prime}},
-\underline{p})$,
and MFRG flows for the SDW and CDW susceptibilities are,
\begin{eqnarray} \label{SDW_susceptibility RGE}
\!\!\!\!\!\frac{d \chi^{\delta}_{\Lambda}(\pi,0)}{d\Lambda}\!\!\!&=&\!
-\!\!\!\int\!\!\!d\underline{p} \frac{d}{d\Lambda}[
G_{\Lambda}(\underline{p})G_{\Lambda}(\underline{p}\! +
\!\underline{Q})](Z^{\delta}_{\Lambda}(\underline{p}))^2, 
\\
\!\!\!\!\!\frac{d Z^{\delta}_{\Lambda}(\underline{p})}{d\Lambda} \!\!\!&=&\!
\!\!\!\!\int\!\!\!d\underline{p}^{\prime}\frac{d}{d\Lambda}[
G_{\Lambda}(\underline{p}^{\prime})
G_{\Lambda}(\underline{p}^{\prime}\!\!+\!\underline{Q})] 
Z^{\delta}_{\Lambda}\!(\underline{p^{\prime}})
g^{\delta}\!(\underline{p^{\prime}}\!,\underline{p}),
\end{eqnarray}
where $\underline{Q} = (\pi, 0)$. For $\delta={\rm SDW}$:
$g^{\delta}(\underline{p^{\prime}},\underline{p})= 
-g(\underline{p}+\underline{Q},\underline{p^{\prime}}, \underline{p})$,
and for $\delta={\rm CDW}$ : 
$g^{\delta}(\underline{p^{\prime}},\underline{p}) =
2g(\underline{p^{\prime}},\underline{p}+\underline{Q},\underline{p}) -
g(\underline{p} +\underline{Q},\underline{p^{\prime}},\underline{p})$.
The function $Z^{\delta}(\underline{p})$ is the effective vertex in the
definition of the susceptibility $\chi^{\delta}$. Its initial RG
value is $1$. The MFRG equations for susceptibilities are solved with
initial condition $\chi^{\delta}_{\Lambda=\Lambda_{0}}=0$.  

In g-ology \cite{Emery,Solyom,Voit} there are only four couplings, 
corresponding to forward ($g_2, g_4$), backward ($g_1$),
and Umklapp ($g_3$), scattering. The charge and the spin parts are
governed by $g_3$  and $g_1$, respectively. Under the MFRG, each one of these couplings
carries frequency dependence, $g_i(\omega_1,\omega_2,\omega_3)$. In the weak e-ph coupling 
limit ($\lambda \ll 1$), the two-step RG is  a good approximation, and the couplings are separated into 
two types: high frequency transfer, $|\omega_1-\omega_4| > \omega_0$, and low frequency transfer,  
$|\omega_1-\omega_4| < \omega_0$. However, our MFRG analysis reveals that the
couplings develop additional non-trivial frequency dependence, particularly when the e-ph coupling is 
comparable to the e-e coupling and $U_{\rm eff} \approx 0$. As we shall see, understanding
this frequency structure is critical to resolving the 
current controversy about the behavior in the region near the CDW-SDW transition. 

\begin{figure}[htb]
\centerline{\includegraphics*[height=0.18\textheight, width=0.24\textwidth,
  viewport=56 240 510 520,clip]{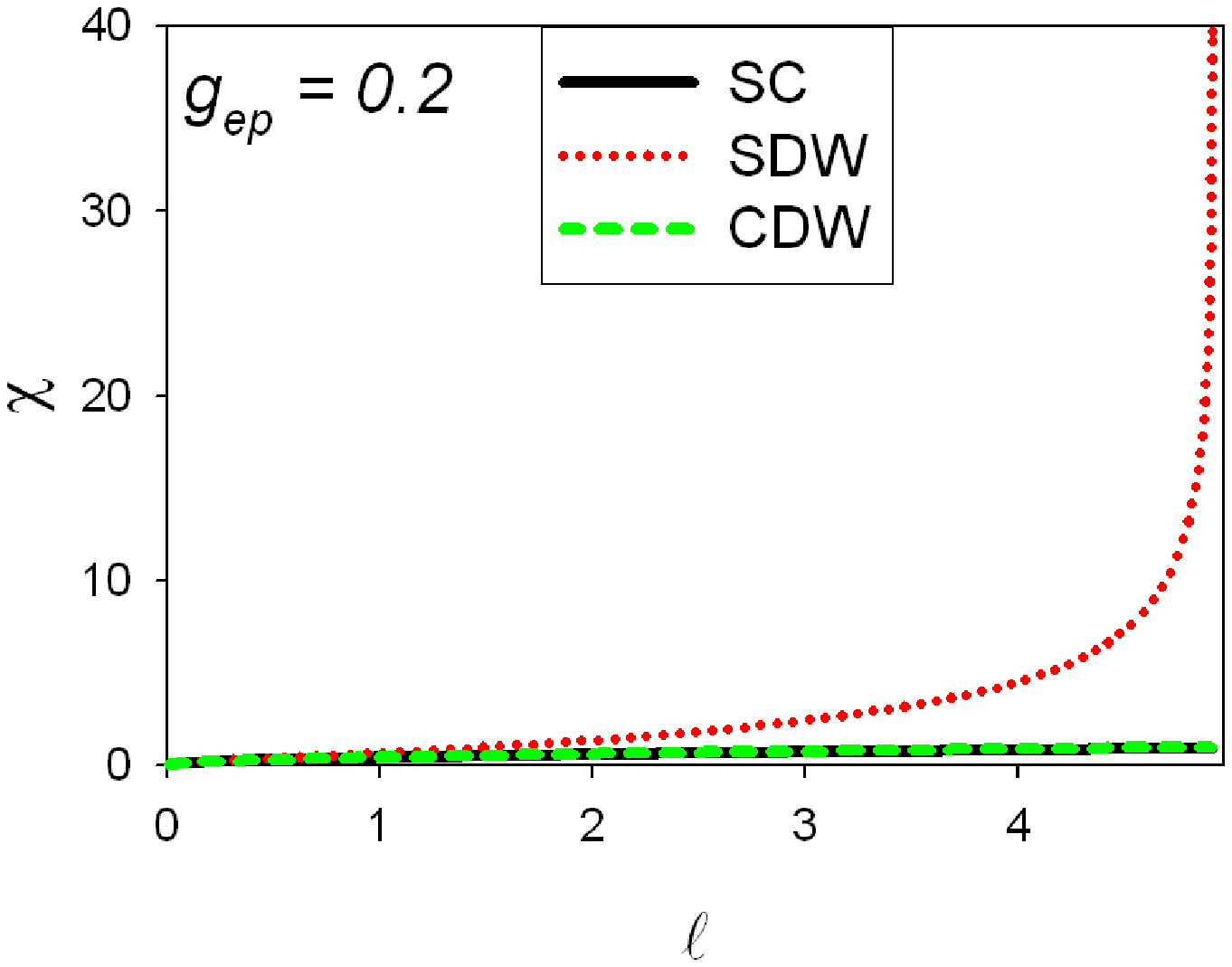} 
\includegraphics*[height=0.18\textheight,width=0.24\textwidth, viewport=56 240
  510 520,clip]{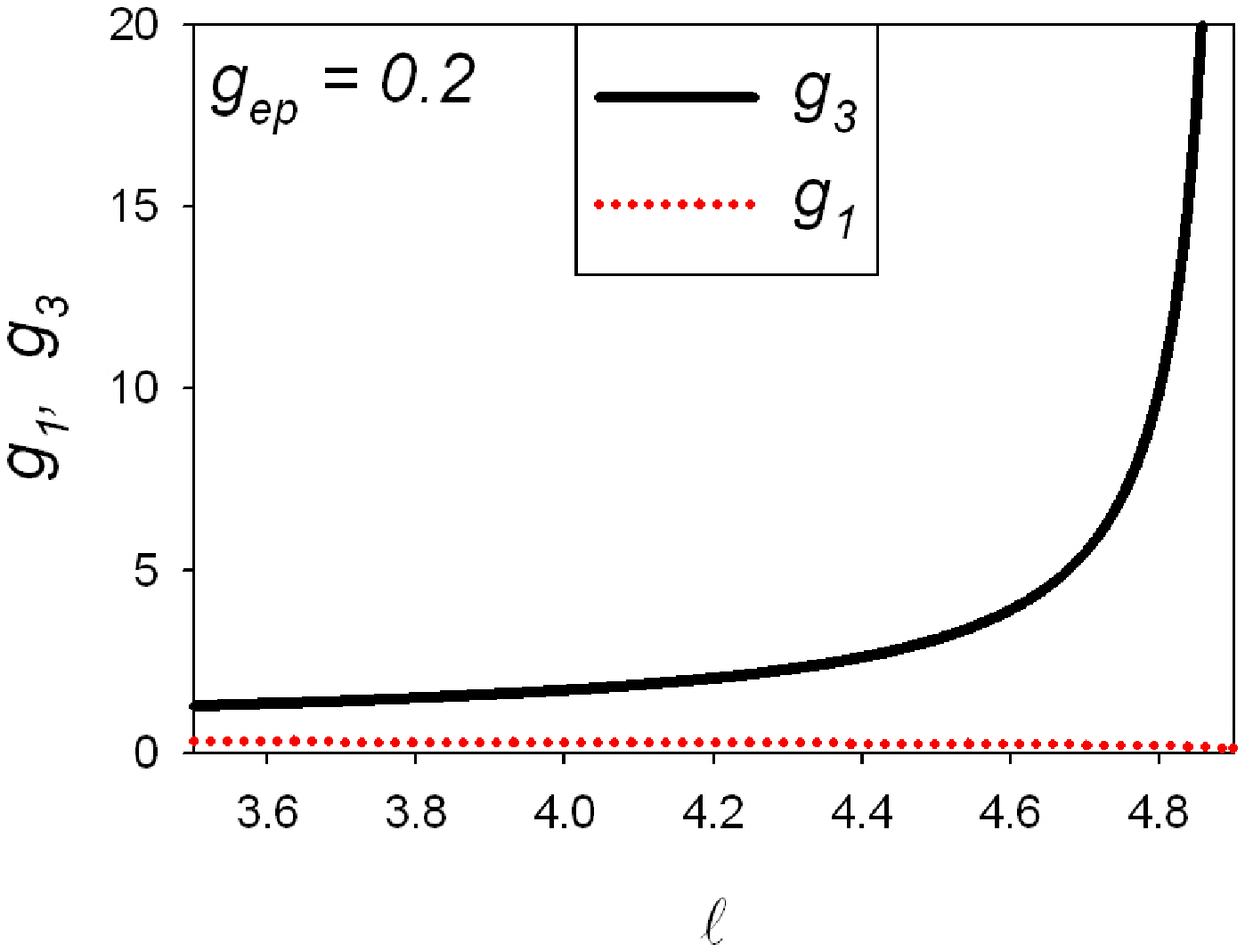}} 
\centerline{\includegraphics*[height=0.18\textheight,width=0.24\textwidth,
  viewport=56 240 510 520,clip]{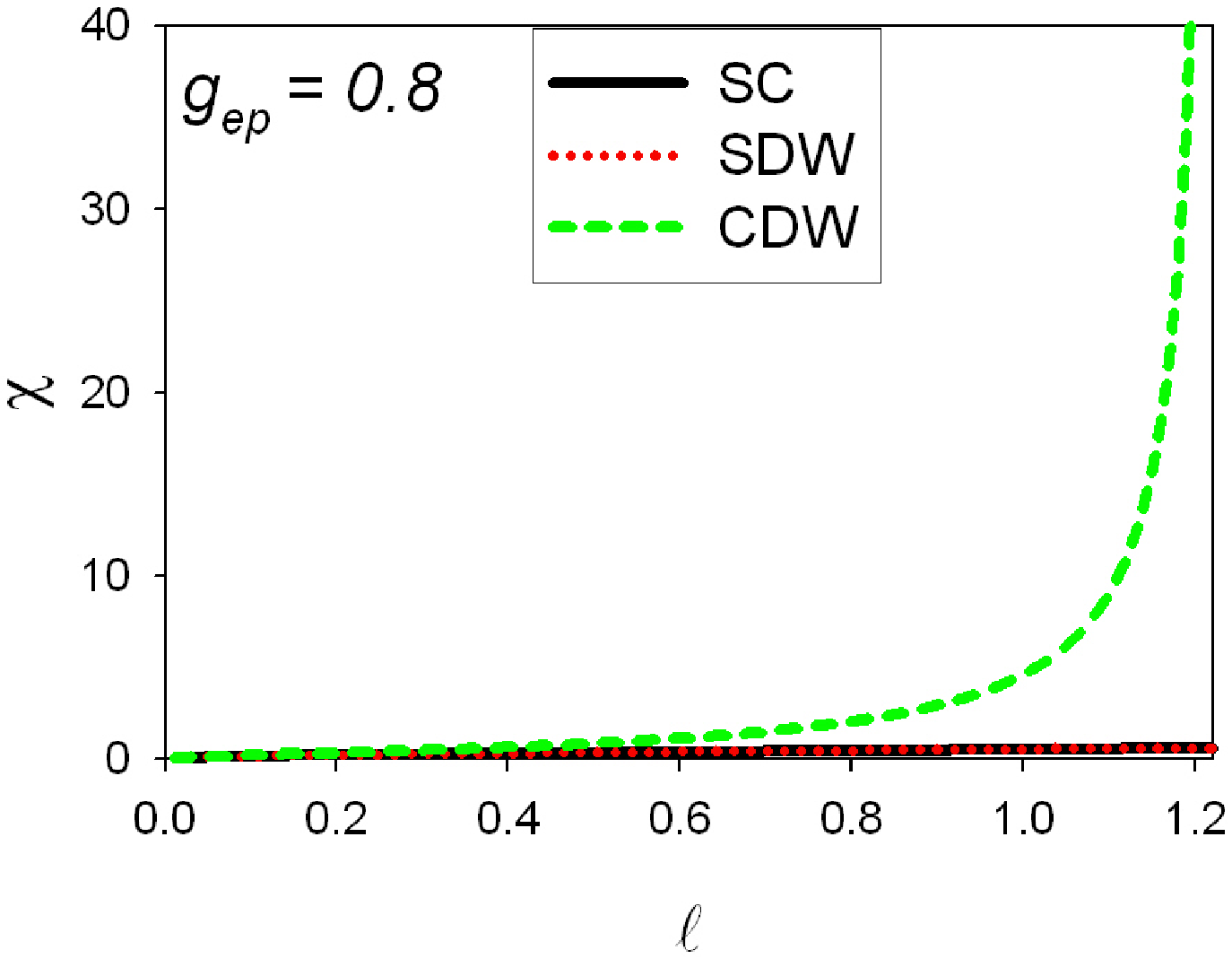} 
\includegraphics*[height=0.18\textheight,width=0.24\textwidth, viewport=56 240
  510 520,clip]{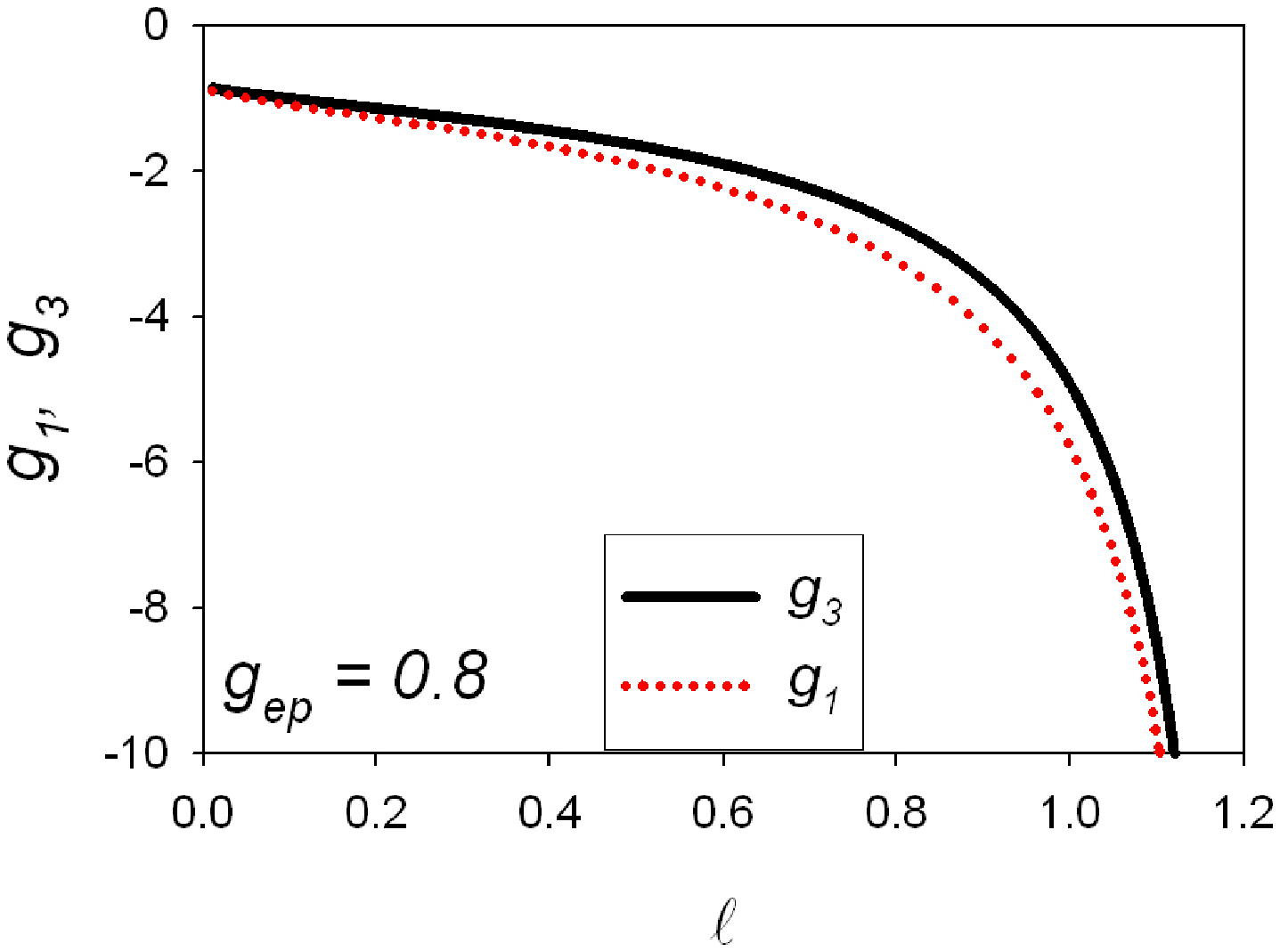}} 
\caption{Left: flows of SC, SDW, and 
CDW susceptibilities for $U=0.5$ and $\omega_{0} = 1.0$. Right: flows of 
Umklapp $g_3$ and back-scattering $g_1$, at zero frequencies. 
Top: $g_{ep} = 0.2$ ($U_{\rm eff} > 0$). Bottom: $g_{ep} = 0.8$ ($U_{\rm eff} < 0)$.}  
\label{HHM:large}
\end{figure}

Deep inside the CDW and SDW regions, we
fix $\omega_0 = 1.0$ and $U = 0.5$, and show results of the RG flows for the susceptibilities and couplings for 
different values of 
%the e-ph coupling 
$g_{\rm ep}$. For small e-ph coupling ($g_{\rm ep} =
0.2$, and  $U_{\rm eff} > 0$), the SDW susceptibility exhibits a strong divergence, while both CDW and SC susceptibilities are 
suppressed (Fig.\ref{HHM:large}, top). This is expected, since the on-site repulsion dominates over the retarded attractive 
interaction mediated by the phonons. A charge gap develops, with no spin gap, which can be inferred from the flow of the 
couplings: Umklapp ($g_3$) diverges, whereas back-scattering ($g_1$) does not.  
For large e-ph coupling ($g_{\rm ep} = 0.8$, and $U_{\rm eff} < 0$), the CDW susceptibility 
diverges (Fig. \ref{HHM:large}, bottom). Now there are both spin and charge gaps, and, correspondingly, both Umklapp ($g_3$) and 
back-scattering ($g_1$) are divergent.  

\begin{figure}[htb]
\centerline{
\includegraphics*[height=0.18\textheight,width=0.24\textwidth, viewport=56 240 510 520,clip]{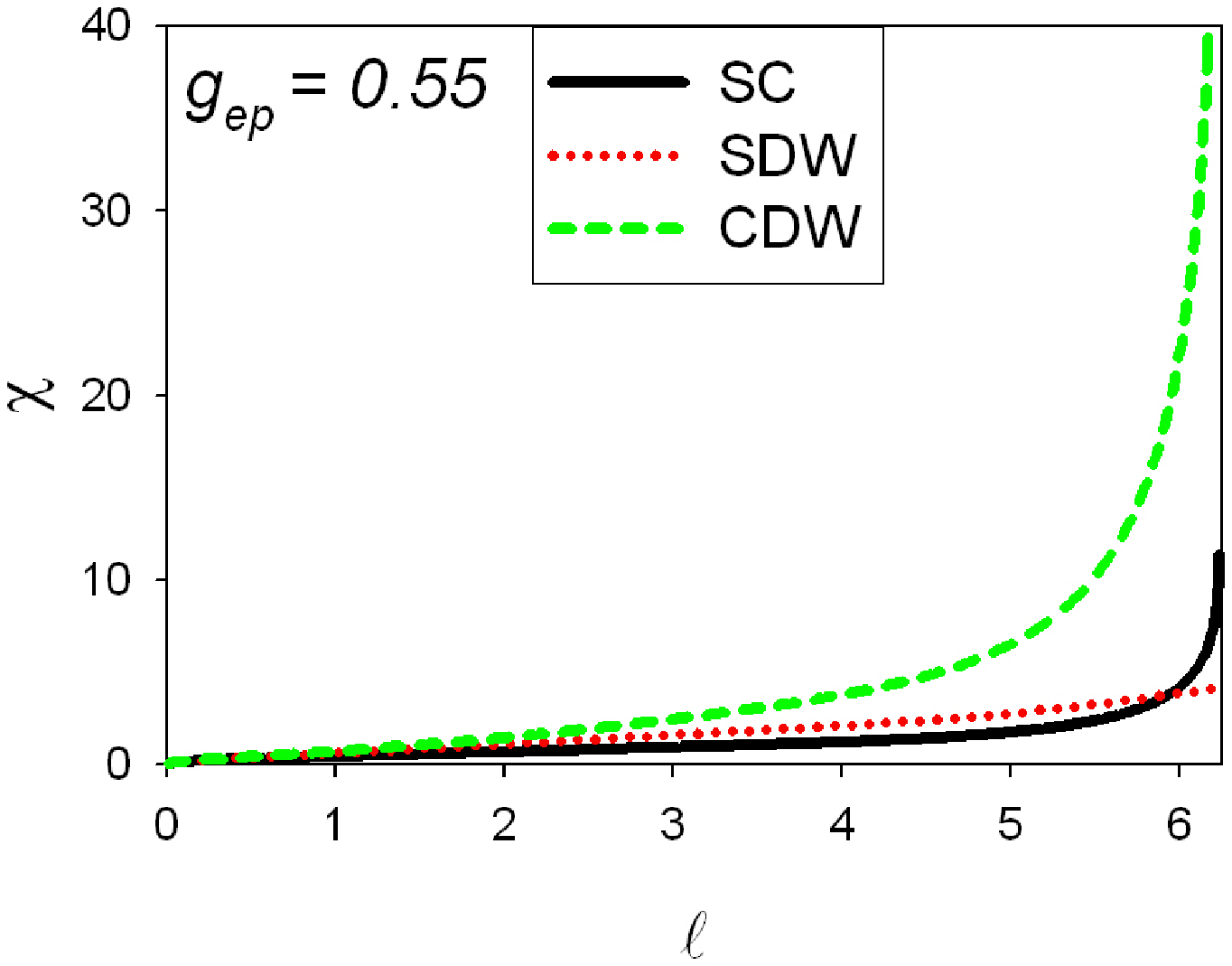} 
\includegraphics*[height=0.18\textheight,width=0.24\textwidth, viewport=56 240 510 520,clip]{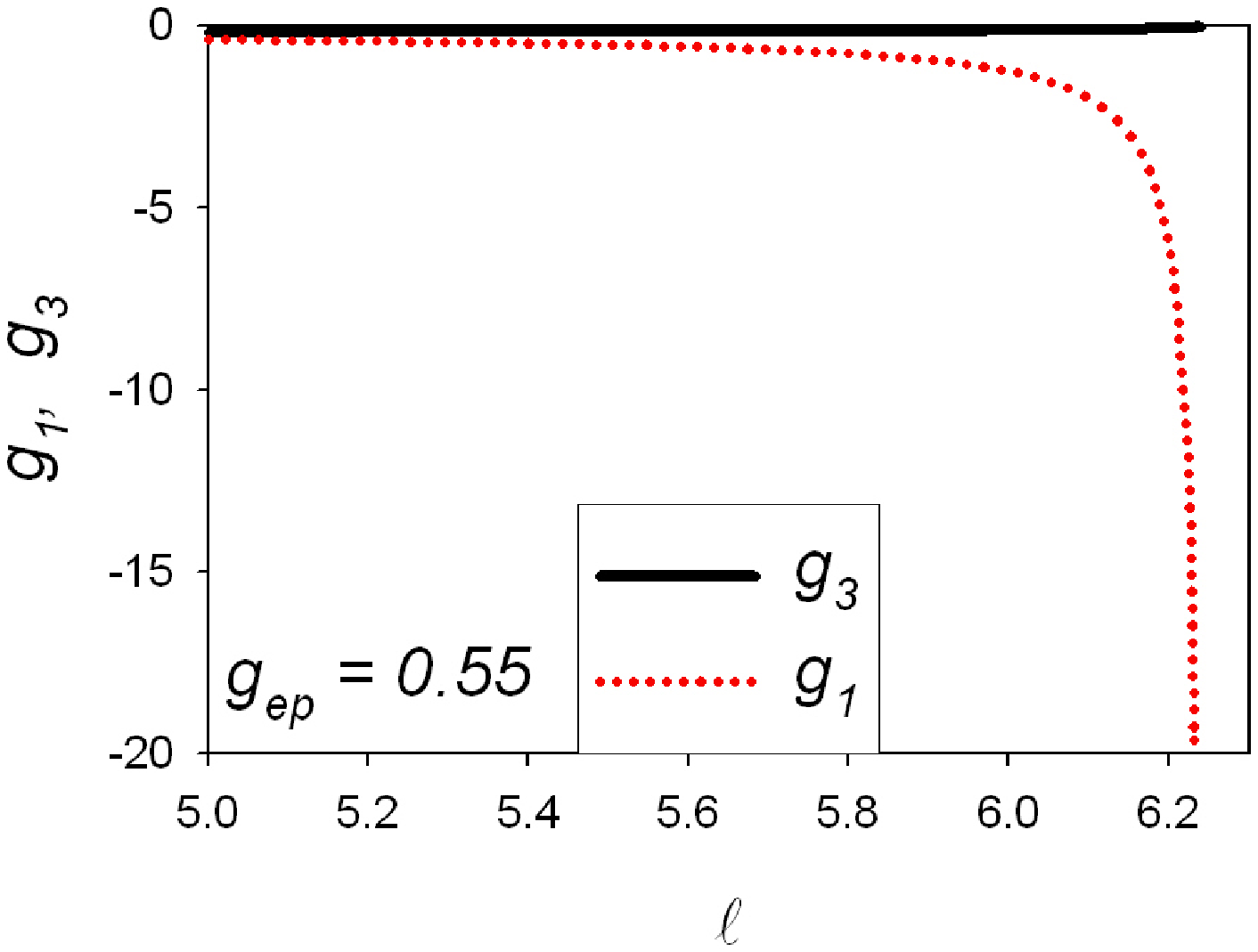}} 
\caption{Left: flow of susceptibilities for
  $U=0.5$, $\omega_{0}=1.0$, $g_{\rm ep}=0.55$ ($U_{\rm eff} < 0$). Right:
  flows of the Umklapp scattering $g_3$ and back-scattering
  $g_1$ at zero frequency.}  
\label{HHM:small}
\end{figure}

We next consider the region close to the CDW-SDW transition where $U_{\rm eff} \simeq 0$. For $U_{\rm eff}$
slightly below zero ($g_{\rm ep} = 0.48$), the behavior of susceptibilities and couplings is qualitatively the same 
as in the rest of the SDW phase (Fig. \ref{HHM:large}, top). The only difference is that the gap decreases and 
eventually goes to zero at the transition. Fig. \ref{HHM:small} shows the flows for $g_{\rm ep} = 0.55$ ($U_{\rm eff}$ slightly  
above zero). The SC susceptibility becomes enhanced, but the CDW susceptibility still dominates. Interestingly,
$g_1(0,0,0)$ diverges but $g_3(0,0,0)$ does not. In 1D problems without retardation, the usual interpretation is that 
the CDW instability occurs when $g_1 \rightarrow - \infty$ and $g_{3} \rightarrow - \infty$
\cite{Emery,Voit,Nakamura}. In the present case, since $g_3(0,0,0)
\rightarrow 0$, we need to look at the frequency dependence of the couplings in order to understand what is 
driving the CDW instability. 
 
\begin{figure}[ht]
\centerline{\includegraphics*[height=0.15\textheight,width=0.24\textwidth,
  viewport=50 100 600 560,clip]{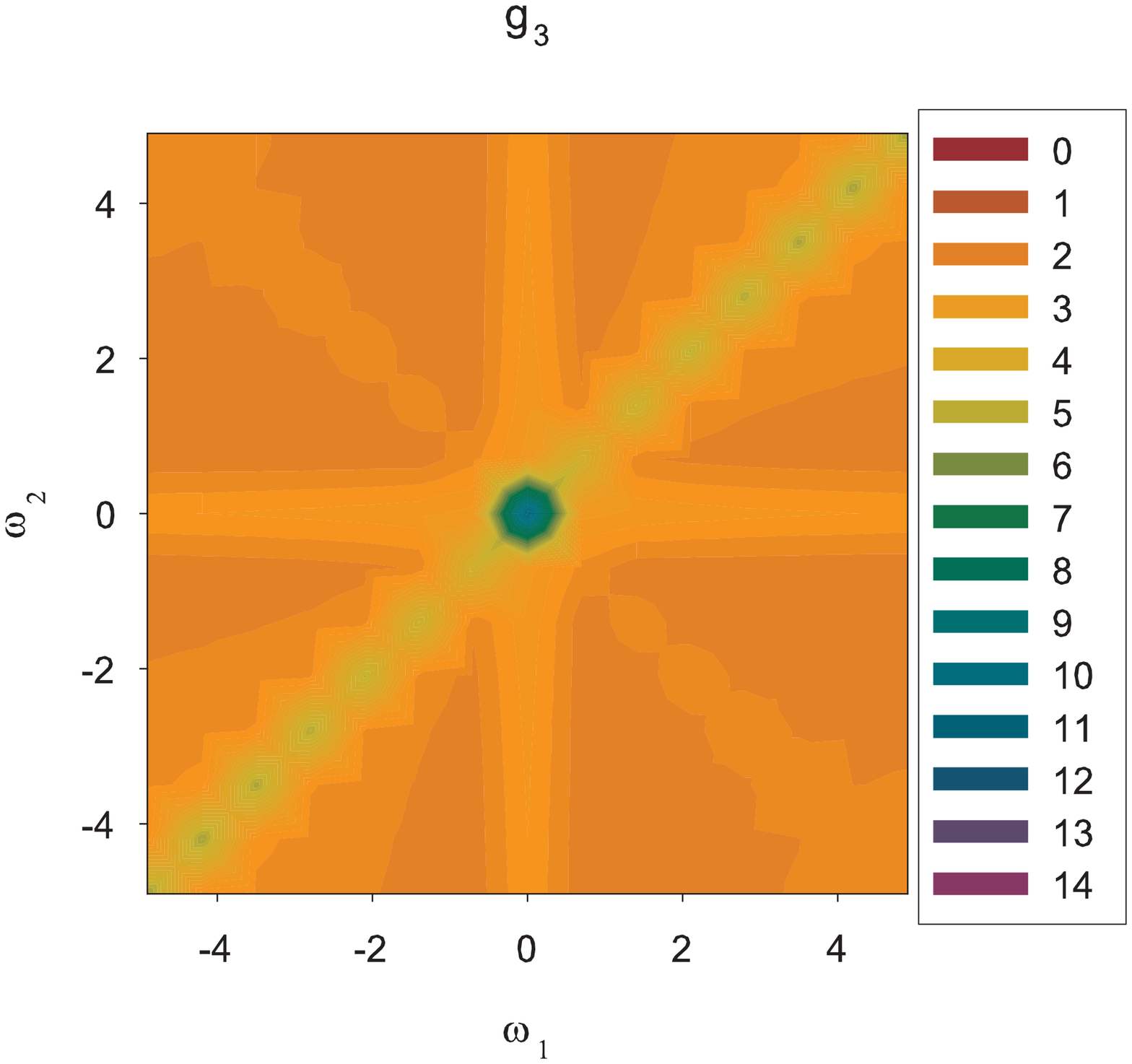} 
\includegraphics*[height=0.15\textheight,width=0.24\textwidth,
  viewport=50 100 600 560,clip]{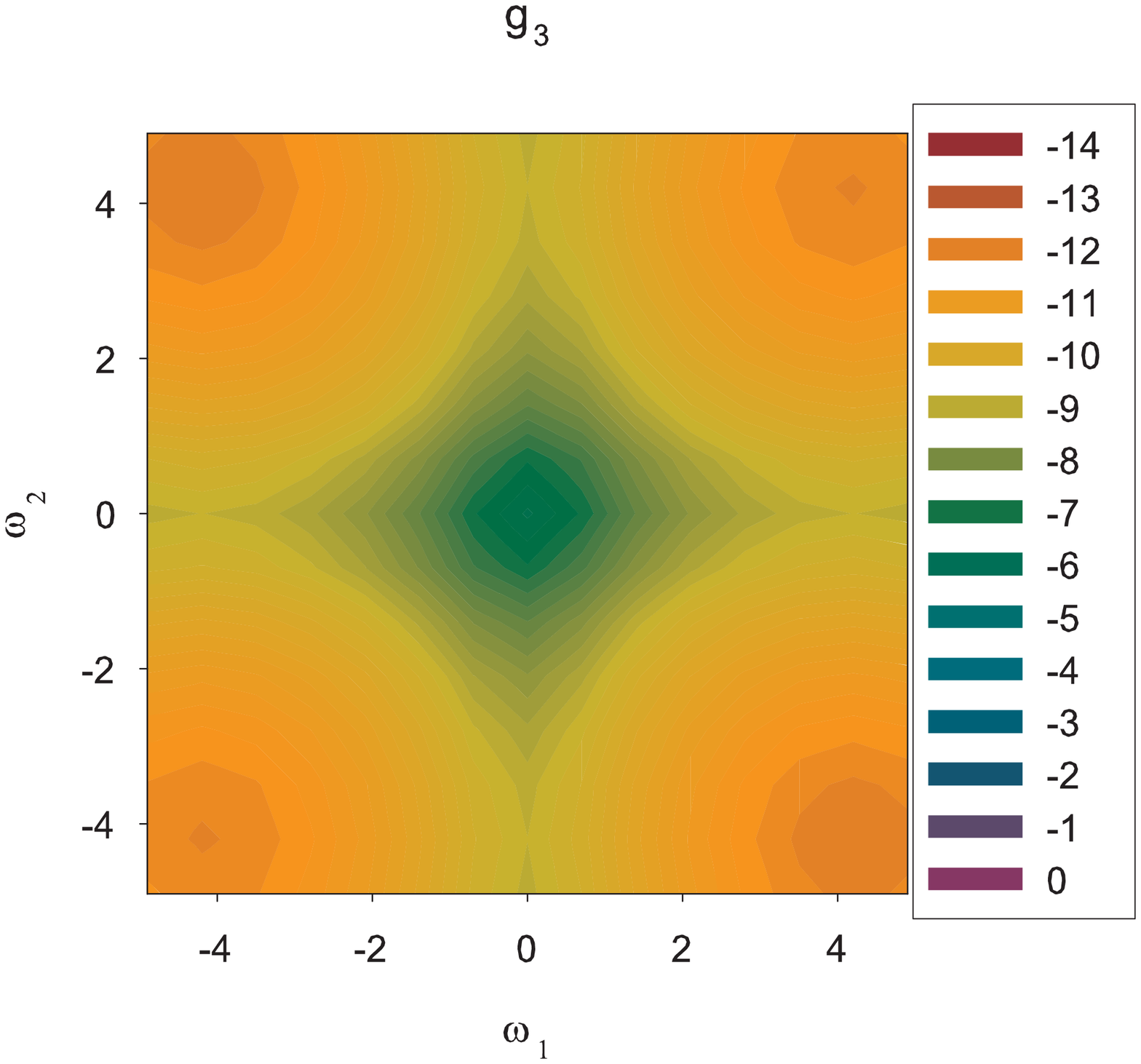} }
\caption{Plots of the Umklapp scattering
  $g_{3}(\omega_1,\omega_2,\omega_2,\omega_1)$ for $U=0.5$, and
  $\omega_{0}=1.0$. Left: $g_{\rm ep} = 0.2$. Right: $g_{\rm ep}
  = 0.8$.}  
\label{fig:g3large}
\end{figure}

In the MFRG approach, we obtain the RG flow of all the
$g_i(\omega_1,\omega_2,\omega_3)$ couplings and self-energies, and therefore can analyze how this frequency dependence 
evolves with the RG flow.  Consider first the cases deep in the SDW and CDW phases.  Fig. \ref{fig:g3large} shows contour 
plots of $g_3(\omega_1,\omega_2,\omega_2,\omega_1)$ which corresponds to an Umklapp process with zero-frequency transfer, 
$|\omega_{1} - \omega_{4}|=0$. We plot the value of the coupling at an RG scale $\ell$ right before the critical 
scale $\ell_c$ when the instability occurs. For the SDW phase (Fig. \ref{fig:g3large}, left), the existence of a charge  
gap is signaled by divergence in the Umklapp channel, and the most divergent $g_3$ couplings are the ones close to zero 
frequency. Deep inside the CDW phase,
$g_3(0,0,0,0)$ also diverges, as we have seen before from
Fig. \ref{HHM:large}. However, the most divergent couplings are for 
large values of $\omega_1$ and $\omega_2$ (see Fig. \ref{fig:g3large}). 
  
\begin{figure}
\centerline{\includegraphics*[height=0.23\textheight,width=0.38\textwidth,
  viewport=50 100 
  600 550,clip]{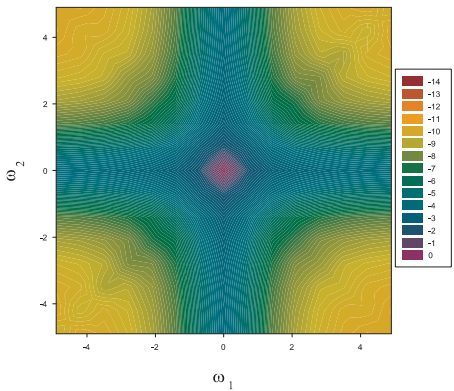}} 
\caption{Plot of the Umklapp scattering
  $g_{3}(\omega_1,\omega_2,\omega_2,\omega_1)$ for $U=0.5$, 
  $\omega_{0}=1.0$, and $g_3=0.55$. Note that $g_3(0,0,0)$ is flowing
  towards zero.}
\label{fig:g3small}
\end{figure}

The situation for $g_{\rm ep}=0.55$, shown in Fig. \ref{fig:g3small}, is more intriguing. Umklapp scattering is renormalized to
large values in most part of the frequency space. However, for frequencies near zero 
Umklapp scattering flows to very small values. From the RG flow of the susceptibilities (Figs. \ref{HHM:large} 
and \ref{HHM:small}), it is clear that there is CDW instability for $U_{\rm eff} > 0$ and a direct transition 
from CDW to SDW. From the frequency dependence of $g_3$ we conclude that close to the transition to the SDW, the CDW 
instability is being driven by Umklapp processes {\it at high frequencies}. These are processes at small frequency 
{\it transfer}, $|\omega_1-\omega_4| \sim 0 < \omega_0$ but that nevertheless involve electrons with 
high frequencies ($\omega_1$ and $\omega_2$). In a two-step RG analysis, the couplings
$g_3(\omega_1,\omega_2,\omega_2,\omega_1)$, with different $\omega_1$ and $\omega_2$ are 
all indistinguishable since $|\omega_1-\omega_4|=0$ for all of them.  Clearly, the two-step RG
fails in this region. 

As an independent (partial) confirmation of our MFRG results, we have also performed determinantal QMC 
\cite{Blankenbecler} calculations for the Holstein model ($U=0$). For the charge exponent,
$K_{\rm CDW}=\lim_{q \rightarrow 0} \pi S^{\rho}(q)/q$, we obtain 
that $K_{\rm CDW} > 1$  when $g_{\rm ep}$ is smaller than some
value that depends on $\omega_0$. This result agrees with that obtained in
\cite{Clay}, using stochastic series expansion QMC \cite{Sandvik}. 
For a Luttinger liquid, the scalings of ground state correlation functions are determined solely by the 
charge ($K_{\rho}$) and spin ($K_{\sigma}$) exponents. For example, in the spin-gapped regime, where 
$K_{\sigma}=0$, CDW and SC correlation functions scale as $O^{{\rm CDW}}(x) \propto x^{-\alpha K_{\rho}} \equiv x^{-K_{{\rm CDW}}}$, 
and  $O^{{\rm SC}}(x) \propto x^{-\beta / K_{\rho}} \equiv x^{-K_{{\rm SC}}}$, with $\alpha=\beta=1$ \cite{Emery,Solyom,Voit}. 
The dominant correlation is of CDW (SC) type for $K_{\rho} < 1$ ($K_{\rho} > 1$). 
This relation is not guaranteed to hold in the presence of phonons and retardation effects \cite{Martin}. 

% In fact, it has been shown 
%that for acoustic phonon in a Luttinger liquid, the parameters $\alpha$ and $\beta$ are different functions of electron 
%charge velocity, phonon velocity and the electron-phonon coupling, so that in general they are not equal to 1 %\cite{Martin}. 
%Therefore, $K_{{\rm CDW}}$ does not have a direct relationship with $K_{{\rm SC}}$ and $K_{{\rm CDW}} > 1$ does not imply
%dominant SC correlations. Explicit functional forms of
%$\alpha$ and $\beta$ have not been calculated for optical phonons, however, a DMRG study of the 1DHHM finds cases where both 
%$K_{{\rm CDW}}$ and $K_{{\rm SC}}$ are larger than one \cite{Tezuka}. 
   
\begin{figure}[thb]
\centerline{
\includegraphics*[height=0.19\textheight,width=0.36\textwidth, viewport=50
220 500 545,clip]{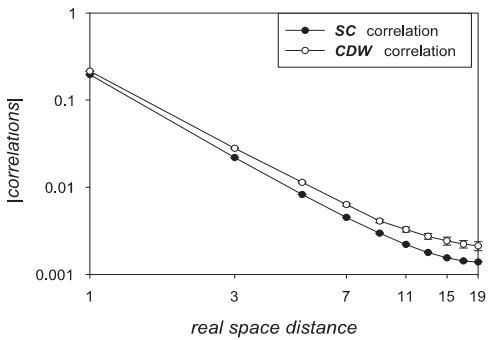}} 
\caption{SC and CDW correlations for 38-sites Holstein model
  ($\omega_{0} = 1.0$, $g_{\rm ep} = 0.5$), with $K_{{\rm CDW}} = 1.032 \pm 0.005 $.}  
\label{HHM:MC_C}
\end{figure}

Using the determinantal QMC allows us to calculate the pairing and charge
correlations directly (Fig. \ref{HHM:MC_C}). We find that the charge
correlation function decays more slowly. This provides, at least for the case $U=0$,
confirmation of our MFRG results and strongly suggests that there is no region of 
dominant SC correlations in the half-filled 1DHHM, even though the scaling exponent of the charge correlation function 
can be larger than $1$.   

In conclusion, we have studied the ground state of 1DHHM at half-filling using the MFRG method. 
This technique enables us to treat retardation 
effects from the phonons in a systematic way. We find SDW and CDW phases, and a direct transition between them. Analysis of the 
frequency dependence of the $g_3$ shows a shift in spectral weight indicating that the CDW instability near the transition is 
driven by dynamical Umklapp processes. Our determinantal QMC results for the charge exponent and correlation functions for the 
Holstein model confirm our MFRG predictions and suggest that having a charge exponent larger than one for finite size system does not mean dominant 
SC correlations because of breakdown of TLL relations due to retardation.

\acknowledgments
We thank Torsten Clay for instructive discussions. A.H.C.N. was
supported through NSF DMR-0343790.


\begin{thebibliography}{99}

\bibitem{Tsai} S.-W. Tsai, A. H. Castro Neto, R. Shankar, D. K. Campbell,
  Phys. Rev. B {\bf 72}, 054531 (2005), Phil. Mag. 86, 2631 (2006). 

\bibitem{Organics}T. Ishiguro and K. Yamaji, {\it Organic Superconductors}
  (Springer-Verlag, Berlin, 1990).

\bibitem{Polymers}{\it Conjugated Conducting Polymers}, edited by H. G. Weiss
  (Springer-Verlag, Berlin, 1992).

\bibitem{Fullerene} O. Gunnarsson, Rev. Mod. Phys. {\bf 69}, 575 (1997).

\bibitem{Holstein} T. Holstein, Ann. Phys. {\bf 8}, 325 (1959).

\bibitem{Hirsch}J.~E.~Hirsch~and~E.~Fradkin, Phys. Rev. B {\bf 27}, 4302
  (1983); J.~E.~Hirsch, Phys. Rev. B {\bf 31}, 6022 (1985).  

\bibitem{Wu} C. Wu, 
%Q. Huang, and X. Sun, 
{\it et al.}, Phys. Rev. B {\bf 52}, R15683 (1995).

\bibitem{Jeckelmann} E. Jeckelmann, C. Zhang, and S. White, Phys. Rev. B {\bf
  60}, 7950 (1999). 

\bibitem{Takada} Y. Takada and A. Chatterjee, Phys. Rev. B {\bf 67}, 081102
  (2003). 

\bibitem{Takada2} Y. Takada, J. Phys. Soc. Jpn. {\bf 65}, 1544 (1996).

\bibitem{Clay} R. T. Clay, R. P. Hardikar, Phys. Rev. Lett. {\bf 95}, 096401
  (2005). 

\bibitem{Tezuka} M. Tezuka, R. Arita, H. Aoki, Physica B {\bf 359}, 708 (2005),
  Phys. Rev. Lett. {\bf 95}, 226401 (2005). 

\bibitem{Feshke} H. Feshke, {\it et al.}, Phys Rev. B {\bf 69}, 165115 (2004).

\bibitem{Bindloss} I. P. Bindloss, Phys. Rev. B {\bf 71}, 205113 (2005).

\bibitem{Shankar}R. Shankar, Rev. Mod. Phys. {\bf 66}, 129 (1994).

\bibitem{Tam} K.-M. Tam, {\it et al.}, cond-mat/0603055.

\bibitem{Emery}V. J. Emery, in {\it Highly Conducting One-Dimensional
    Solids}, p. 327, edited by J. T. Devreese, 
%R. Evrand, and V. van Doren
 {\it et al.}   (Plenum, New York, 1979).

\bibitem{Solyom}J. S\'{o}lyom, Adv. Phys. {\bf 28}, 201 (1979).

\bibitem{Voit} J. Voit, Rep. Prog. Phys. {\bf 58}, 977 (1995).

\bibitem{Nakamura} M. Nakamura, Phys. Rev. B {\bf 61}, 16377 (2000).

\bibitem{Martin} D. Loss and T. Martin, Phys. Rev. B {\bf 50}, 12160 (1994).

\bibitem{Blankenbecler} R. Blankenbecler, 
%D. J. Scalapino, and R. L. Sugar,
{\it et al.}, Phys. Rev. D {\bf 24}, 2278 (1981). 

\bibitem{Sandvik} O. F. Syljuasen and A. W. Sandvik, Phys. Rev. E {\bf 66},
  046701 (2002). 


\end{thebibliography}
\end{document}